\documentclass[12pt]{article}
\usepackage{amsmath,amssymb,amsfonts,epsfig}
\usepackage[nosort]{cite}
\usepackage{bm}

\makeatletter\@addtoreset{equation}{section}\makeatother

\def\XXint#1#2#3{{\setbox0=\hbox{$#1{#2#3}{\int}$}
\vcenter{\hbox{$#2#3$}}\kern-.5\wd0}}

\allowdisplaybreaks

\input epsf

\begin{document}

\thispagestyle{empty}
\vspace*{-2em}
\begin{flushright}
RIKEN-TH-206\\ 
February 2011
\end{flushright}
\vspace{0.3cm}
\begin{center}
\Large {\bf
Deformation of half-BPS solution in ABJM model \\ 
and instability of supermembrane}

\vspace{0.7cm}

\normalsize
 \vspace{0.4cm}

Tsunehide {\sc Kuroki}$^{a,}$\footnote{e-mail address:\ \ 
{\tt tkuroki@rikkyo.ac.jp}}
,
Akitsugu {\sc Miwa}$^{b,}$\footnote{e-mail address:\ \ 
{\tt amiwa@riken.jp}}
and 
Satoshi {\sc Okuda}$^{a,}$\footnote{e-mail address:\ \ 
{\tt okudas@rikkyo.ac.jp}}

\vspace{0.7cm}

$^a$
{\it 
Department of Physics, Rikkyo University, 
Tokyo 171-8501, Japan}\\

\vspace{0.4cm}

$^b$
{\it Theoretical Physics Lab., RIKEN Nishina Center,  Saitama 351-0198, Japan}\\

\vspace{1cm}
{\bf Abstract}
\end{center}
It is well-known that a supermembrane in the light-cone gauge has 
a continuous spectrum and is unstable. Physical interpretation of this instability 
is that a supermembrane can have a long thin tube without cost of energy and 
consequently it becomes a spiky configuration in which multiple membranes are connected 
by thin tubes.

On the other hand, the ABJM model was proposed as a low-energy description 
of multiple M2-branes in the static gauge. It is natural 
that an M2-brane is also unstable in this gauge if we believe the physical picture 
in the light-cone gauge. In order to examine this, we construct a BPS solution explicitly
both in the Nambu-Goto action of a supermembrane in the static gauge and 
in the ${\rm U}(1)\times {\rm U}(1)$ ABJM  model, which represents intersecting M2-branes. 
Since this configuration is regarded as a single M2-brane emitting another one, 
we study the instability of an M2-brane by analyzing fluctuations around it. 
We show that a zero mode exists which can deform the configuration. 
For comparison, we also examine a similar configuration 
on the D2-brane and check that it does not have such zero modes 
under a fixed string charge.
Furthermore we confirm that the novel Higgs mechanism translates our BPS solution 
in the ABJM model into that in the D2-brane world volume theory, 
where the winding number of the former around the fixed point of the orbifold 
becomes the number of strings ending on the D2-brane in the latter.

\newpage
\setcounter{page}{1}

\tableofcontents

\setcounter{footnote}{0}

\section{Introduction}
\label{sec:intro}

In nonperturbative formulation of string theory, 
identification of fundamental degrees of freedom is one of the most important problems. 
M-theory \cite{Witten:1995ex} defined in eleven dimensions is proposed 
as a unifying framework of string theory, where a membrane (an M2-brane) is believed 
to play an essential role like a fundamental string in string theory. 
In fact,  if the eleventh dimension is compactified on a circle, an M2-brane 
wrapped in this direction is regarded as a fundamental string in ten dimensions. 
Therefore it is natural to expect 
that M-theory should be formulated in terms of a supermembrane as a fundamental object. 

However, it is well-known that a quantum supermembrane theory has a serious problem 
that it has a continuous spectrum \cite{deWit:1988ct}.
This means that it is unstable quantum mechanically. 
More precisely, if we formulate a supermembrane theory in eleven dimensions 
in the light-cone gauge, the action is reduced to a matrix quantum mechanics 
with the supersymmetric Yang-Mills type action, 
where the gauge group is the area preserving diffeomorphism on the world volume
\cite{deWit:1988ig}. 
Since this action has eight scalar fields and the scalar potential is given 
in terms of their commutators, there are flat directions 
where the scalar fields have arbitrary large values 
as long as they are diagonal. It was shown rigorously in \cite{deWit:1988ct} that 
the presence of these flat directions makes a spectrum of supermembrane continuous. 
This fact implies that a supermembrane is pathological quantum mechanically. 
As explained in \cite{deWit:1988ct}, the physics behind this instability is as follows: 
a supermembrane has mass proportional to its 2-dimensional area. 
Hence it can have a long tube (spike) without cost of energy 
as long as it is sufficiently thin. 
Thus a supermembrane can emit a tube which becomes thinner and longer, 
and eventually by an entropic effect, a supermembrane becomes 
a spiky configuration in which multiple supermembranes are connected with each other 
by thin tubes. Namely, quantum mechanically a single supermembrane does not make sense 
and it is a multi-body problem in nature, which can be regarded as the origin 
of the continuous spectrum. 

On the other hand, some years ago a low energy effective action of multiple M2-branes 
was proposed \cite{Aharony:2008ug}. It is defined as three-dimensional 
${\cal N}=6$ superconformal ${\rm U}(N)\times {\rm U}(N)$ 
Chern-Simons theory with level $(k,-k)$. 
It is conjectured to describe $N$ M2-branes located at the fixed point 
of the $\bm C^4/\bm Z_k$ orbifold in the static gauge. 
If this conjecture is true, one of the most important issues would be 
whether the ABJM model resolves the instability of a supermembrane shown in \cite{deWit:1988ct}. Namely, if we believe the physical picture mentioned above behind the instability, 
the supermembrane should be also unstable in other gauges 
than the light-cone gauge. From this point of view, it is natural to expect that 
the ${\rm U}(1) \times {\rm U}(1)$ ABJM model would show the instability
of a single M2-brane as well. 
Moreover the ABJM model is formulated in the background
which is different from the flat space. 
Thus it is interesting to examine the instability
of M2-brane in the context of the ABJM model.

Our idea is quite simple: we first construct a classical configuration 
both in the Nambu-Goto action of a supermembrane
and in the ${\rm U}(1)\times {\rm U}(1)$ ABJM model, 
which represents a single M2-brane emitting another M2-brane. 
Fortunately it preserves a half of supersymmetries in both models. 
This BPS solution has a scale corresponding to the size of the configuration, 
thus it breaks the conformal symmetry in the ABJM model explicitly. 
Then we construct quadratic Lagrangian for fluctuations around it. 
If this Lagrangian has a flat direction which is associated with 
the deformation of the original configuration, 
it would be a sign of instability of the classical configuration itself 
against a small fluctuation. 
It is however worth noticing that such a perturbative result of the flat direction
is not direct evidence of instability, but only its sign. 
A good example is provided by a bosonic membrane, 
where it has a flat direction and thus looks unstable at first sight, 
but it is known to be stable \cite{deWit:1988ct}.    
Therefore even if there exists a flat direction in the fluctuation Lagrangian, 
it would need careful analysis to confirm  that a supermembrane is indeed unstable 
due to it by an entropic effect. 
Notice that in the present case we should not have
tachyonic fluctuations because we are considering a BPS solution. 
Then instead of looking for a flat direction of the fluctuation Lagrangian explicitly, 
we construct a Hamiltonian from the Lagrangian and examine 
whether it has zero modes in both Nambu-Goto and ABJM. 
In both models the Hamiltonian is actually represented as  
a sum of squares, which makes identification of zero modes easier, 
where in the ABJM we do not have to take any gauge of 
${\rm U}(1)\times {\rm U}(1)$ gauge symmetry. 
In fact, if we substitute the zero mode found in this way 
into the fluctuation Lagrangian, it vanishes up to boundary contributions. 
Thus the zero mode would give rise to a flat direction
at least in a perturbative sense.  
Hence their existence can be considered as one of signs of 
instability of the membrane. Moreover, in the ABJM model 
the zero mode we identify corresponds to a fluctuation 
which makes the configuration thin in accordance with the physical picture 
behind the instability of a supermembrane in the light-cone gauge mentioned above. 
In the ABJM model, the existence of such a zero mode 
is guaranteed by the fact that 
the BPS solution breaks the conformal symmetry.
Here it should be noticed that as far as zero modes are concerned, 
the ABJM model should incorporate them correctly 
even if it is a low-energy effective theory of an M2-brane. 
We further confirm that the situation does not change 
in a nonlinear extension of the ABJM model proposed recently in \cite{Sasaki:2009ij}, 
thus higher derivative corrections do not stabilize an M2-brane. 

For comparison and in order to check the validity of our analysis, 
we also consider a spike solution which is again half-BPS 
in the D2-brane world volume theory \cite{Callan:1997kz, Gauntlett:1997ss}. 
Since it is regarded as a fundamental string ending on the D2-brane 
which has mass proportional to its length, the spectrum of fluctuations\footnote
{Relations between fluctuations around the spike solution given in \cite{Callan:1997kz} 
and the boundary conditions the D-brane imposes on a string 
are discussed e.g. in \cite{Callan:1997kz,Rey:1998ik,Savvidy:1999wx}. }
around it should not contain a zero mode 
corresponding to its deformation, which we check in the same way as 
in the ABJM case.
In the D2-brane case, we find that a zero mode is not allowed 
if the string charge is fixed.
On the other hand, in the supermembrane case, the zero mode exists 
even if we fix its winding number around the fixed point of the orbifold. 
Thus they show a clear difference in the existence of a zero mode 
under fixed corresponding charges. 
We regard such a sharp contrast as a manifestation of a difference 
in stability of a membrane and string. 

We also clarify how both BPS solutions 
in the ABJM model and in the D2-brane world volume theory are connected 
by the novel Higgs mechanism 
\cite{Mukhi:2008ux,Ho:2008ei,Honma:2008jd,Pang:2008hw}.\footnote
{Correspondence at the level of the half-BPS equations in both sides is demonstrated 
in \cite{Fujimori:2010ec}.} 
It is explicitly shown that the winding number of the BPS
solution in the ABJM model around the fixed point of the 
orbifold are translated into the number of strings 
ending on the D2-brane in accordance with the M-theory interpretation. 
In fact, this observation plays a crucial role in the discussion of stability 
of a membrane and a string as mentioned above. 
  
This paper is organized as follows: in the next section we consider the Nambu-Goto action 
of a supermembrane and construct a half-BPS solution. Then we show that 
fluctuations around it contain a zero mode which deforms the configuration 
in the case of the noncompact target space, but that this is not the case 
with the compact one. 
In section 3 we turn to the ABJM model and repeat the same argument to show that 
the half-BPS configuration also has a zero mode 
which indicates instability of an M2-brane 
perturbatively. In section 4 we construct a spike solution on a D2-brane 
which is identified as a fundamental string ending on the D2-brane, 
and show that it does not contain a zero mode 
if the string charge is fixed,
which makes manifest a difference in stability of an M2-brane and a string. 
The final section is devoted to conclusions and discussions. 
In appendix \ref{SUSY-D2} we give a half-BPS condition in the Dirac-Born-Infeld action 
for a D2-brane. 
In appendix \ref{app:bc} 
we discuss possible boundary contributions to the energy of the BPS solution.

\section{Deformation of the Nambu-Goto supermembrane}

Based on the physical interpretation for the instability
of a supermembrane, we construct a BPS configuration 
in the Nambu-Goto action, and examine whether it allows 
deformations which make the configuration thinner.
We start with a supermembrane in the flat 
noncompact 11-dimensional spacetime.
Next we compactify the 11th direction on an ${\rm S}^1$
and study a supermembrane winding around it.

\subsection{Noncompact target space}

The bosonic part of the action for a single supermembrane 
on the flat 11-dimensional spacetime is given by 
the Nambu-Goto action
\begin{align}
S_{\rm M2} 
& =
-
T_{\rm M2}
\int d^3 x 
\sqrt{- \det g_{\mu \nu} } \nonumber \\
& =
- 
T_{\rm M2} 
\int d^3 x
\sqrt{
- \det 
\big(
\eta_{MN}^{(11)}\partial_\mu X^M \partial_\nu X^N
\big)
} \label{NG}\\
& =
-T_{\rm M2}
\int d^3 x
\sqrt{
- \det 
\big\{
\eta_{\mu \nu} 
+ 
{(2 \pi T_{{\rm M}2})}^{-1}
\big( 
\partial_\mu Y^A \partial_\nu Y_A
+
\partial_\nu Y^A \partial_\mu Y_A
\big)
\big\}
}\,,
\end{align}
where each index takes values as
$\mu,\nu = 0 \sim 2$,
$M,N = 0 \sim 9, 11 $
and $A = 1 \sim 4$.
$ x^\mu $ are the world volume coordinates
and $ X^M $ are the target ones.
$\eta^{(11)}_{MN}$ and $\eta_{\mu \nu}$
are the flat Minkowski metric, ${\rm diag}(-1,1,1,\cdots)$.
In the third line we took the static gauge,
$x^\mu = X^\mu$,
and introduced the complex coordinates as
\begin{align}
& 
Y^1=\sqrt{\pi T_{{\rm M}2}}(X^3 + i X^4), \quad 
Y^2=\sqrt{\pi T_{{\rm M}2}}(X^5 + i X^6), \\
& 
Y^3=\sqrt{\pi T_{{\rm M}2}}(X^7 + i X^8), \quad
Y^4=\sqrt{\pi T_{{\rm M}2}}(X^9 + i X^{11}).
\end{align}
$Y_A$ are the complex conjugates of $Y^A$.

Since the supersymmetry plays a crucial role
in the analysis of \cite{deWit:1988ct}, 
we consider a configuration which 
preserves some supersymmetries.
The residual supersymmetries preserved 
by the membrane on the flat background 
is given by \cite{Bergshoeff:1987dh}
\begin{equation}
\Gamma \epsilon = \epsilon, \quad 
\Gamma = {1 \over 3! \sqrt{-\det g_{\mu \nu}}} 
\epsilon^{\mu \nu \rho}
\partial_\mu X^M 
\partial_\nu X^N 
\partial_\rho X^P
\Gamma_{MNP}.
\end{equation}
Here $\Gamma_M$ are the 11-dimensional Dirac matrices
which satisfy the relation 
$\{\Gamma_M, \Gamma_N\} = 2 \eta^{(11)}_{MN}$\,,
and $\epsilon$ is a constant
32-component Majorana spinor.
We consider the static configuration
$\partial_0 Y^A =0$ 
and further assume $Y^{1,2,3} = 0$\,.
Then a holomorphic configuration $Y^4 = w(z)$, with $z = x^1 + i x^2$, 
or an anti-holomorphic one $Y^4= \bar w (\bar z)$
preserves the quarter of the 32 supersymmetries 
which are specified by 
$(1-\Gamma_{012}) \epsilon = (1 - \Gamma_{09(11)})\epsilon=0$
or 
$(1-\Gamma_{012}) \epsilon = (1 + \Gamma_{09(11)})\epsilon=0$,
respectively. From the point of view of the world volume theory 
it is called the half-BPS.
In the following we study the holomorphic case.

The holomorphic configuration $Y^4 = w(z)$ can be specified 
by the condition $f(w,z)=0$ with some holomorphic
function $f(w,z)$.
We want a configuration which asymptotically approaches 
a single flat membrane extended in the $(X^1,X^2)$-plane
in the limit $|z| \to \infty$,
while which is deformed around the origin.
Based on this viewpoint, we assume that there is 
a unique zero of $f(w,z)$ for any fixed value of $z$.
We also assume that the membrane does not go through
the same point on the target space twice,
namely we assume that for a fixed $w$,
$f(w,z)$ has, if any, a unique zero.
These requirements restrict the function $f(w,z)$ 
to the form $f(w,z) = a wz + b w + c z + d$.
The condition that the configuration is reduced to 
a flat membrane in the $(X^1,X^2)$-plane in the limit $|z| \to \infty$
also imposes $a \neq 0$.
Then by using the translation along 
the world volume and the target space,
we take $b=c=0$.
The rotation symmetry on the $(X^9, X^{11})$-plane 
can be used to set the relative phase 
between $a$ and $d$ to be zero.
After these considerations, 
the resulting configuration is given by
\begin{equation}
Y^4 = w(z) = {\alpha \over z}\,,
\label{a/z}
\end{equation}
where $\alpha$ is a real number.\footnote{
In \cite{Hashimoto:1997px} this kind of analysis is applied 
to derive a different BPS configuration.}

Next we study fluctuations around the configuration.
We consider that of the field $Y^4$, 
assuming the rest to be zero, $Y^{1,2,3}=0$ for simplicity.
For later convenience, 
we consider the fluctuation around 
a generic holomorphic configuration 
$Y^4 = Y_{\rm cl}(z)$ as 
$Y^4 = Y_{\rm cl}(z) + Y(x^0,z, \bar z)$.
By keeping only the quadratic terms, which is valid for small fluctuations, 
the action for the fluctuation becomes 
\begin{align}
S^{(2)}_{\rm M2}
=
- {1 \over 2 \pi}\int d^3 x
\bigg(
-|\dot Y|^2
+2|\bar\partial Y|^2
+
2
{\pi T_{\rm M2}-|\partial Y_{\rm cl}|^2  
\over 
\pi T_{{\rm M}2} + |\partial Y_{\rm cl}|^2}
\big|
\partial Y 
\big|^2
\bigg)\,.
\label{S(2)}
\end{align}
As explained in the introduction, we study a zero mode
of the Hamiltonian since it would give rise to a flat direction, 
and can be a sign for the instability of the solution \eqref{a/z}.

The conjugate momenta are given by
\begin{equation}
P_Y = {1 \over 2 \pi} \dot Y^\dag, \quad
P_{Y^\dag} = {1 \over 2 \pi} \dot Y,
\end{equation}
and the Hamiltonian following from \eqref{S(2)} is 
\begin{align}
{\cal H} 
& =
2 \pi P_Y P_{Y^\dag}
+ 
{2T_{\rm M2} \over \pi T_{\rm M2} +
|\partial Y_{\rm cl}|^2}
|\bar \partial Y|^2\,.
\label{H-NG}
\end{align}
In the final expression \eqref{H-NG},
we neglected total derivatives. 
However, this does not necessarily mean that they actually vanish, 
because we will allow a singular fluctuation $Y$ as shown in a moment.
In order to take account of total derivatives for a singular configuration, 
we would have ambiguities associated with regularizations 
and possible boundary terms.  
Hence we concentrate on the bulk contribution to the Hamiltonian. 
For some concrete consideration of boundary terms, see appendix \ref{app:bc}.
\eqref{H-NG} shows that the quadratic Hamiltonian vanishes
for any static holomorphic fluctuation $Y = Y(z)$.
Hence it can be a candidate for the marginal deformation.
We further argue that possible deformations should 
be no more singular than the original configuration
$Y_{\rm cl}(z)$.\footnote
{Another criterion for allowed fluctuations around a singular 
solution is found, e.g. in \cite{Yoneya:1977yi}.}
In the case of the configuration 
$Y_{\rm cl}(z) = \alpha/z$, 
these requirements allow only
the fluctuation of the overall factor, 
namely $Y^4 =  (\alpha + \delta \alpha)/z$.
Then it is a candidate for the marginal deformation 
which destabilize the original configuration.\footnote{
Notice that the zero modes originating from the
global symmetries like translations and rotations 
which are broken by the configuration $Y=Y_{\rm cl}(z)$ 
are also given by the holomorphic functions.
Some of these are more singular than $Y_{\rm cl}$
at the origin and others are not.
In any case we expect that these are not 
related to the instability of the supermembrane.
}

Let us clarify the relation between 
the above analysis and the stability analysis 
of \cite{deWit:1988ct}.
First of all the solution \eqref{a/z}
is not like a ``spike'' in the sense
that in a small $|z|$ region it becomes 
a flat 2-dimensional plane extended in the transverse 
$X^9$, $X^{11}$ directions.
In the above analysis, 
the emphasis is placed not 
on the shape of the classical configuration
but on the existence of its marginal deformation. 
In fact, the fluctuation of the overall constant
do change the form of the membrane into the one
which is thinner compared to the original one. 
\eqref{a/z} shows that under fixed $|Y^4|$  
if we reduce $\alpha$, so does $|z|$. 
This means that the configuration becomes thinner 
by the change of the overall constant. 
This is a reminiscent of the marginal deformation
discussed in \cite{deWit:1988ct}. 
Notice that we can describe such a deformation through $Y^4$ 
as a result of the static gauge. 
In the analysis based on the ABJM model, in the following section,
a solution similar to \eqref{a/z} will be discussed, 
which represents a spike-like configuration.
Also one should notice that the existence
of the marginal deformation in the above analysis
does not immediately mean the instability. 
In order to judge whether the configuration
is stable or not, we need to go beyond perturbative analysis 
and see whether or not the flat direction is lifted by quantum effect. 
Such an interesting and important analysis is 
beyond the scope of the present paper;
we concentrate on the existence of the flat direction
at the quadratic level.

\subsection{Compact target space}\label{Compact target space}
Next we consider the case
where the 11th direction $X^{11}$
is compactified on a circle and 
a supermembrane is wrapped around it.
In this case, since the compact direction has 
a finite size, the membrane can not shrink 
to an arbitrary thin configuration.
Then one may expect that there are 
no marginal deformations.
We will see how this intuitive 
expectation is realized.

Let us compactify the 11th direction
with the period $2 \pi R$, 
$X^{11} \sim X^{11} + 2 \pi R$.
Then a holomorphic configuration $Y^4=w(z)$
which is consistent with the compactification 
is specified by introducing the variable 
$t = \exp( {w \over \sqrt{\pi T_{\rm M2}}R })$ and require the 
condition $f(t,z) = 0$. 
We consider a membrane solution which
is winding once around the ${\rm S}^1$ direction.
This requires the function $f(t,z)$ to be linear in $z$.
Also we assume that there is only a unique zero of $f(t,z)$ 
for a fixed value of $z$, which seems to be
natural for the spike-like configuration.
Then again we have $f(t,z) = a tz + b t + c z + d$.
We assume that $|z| \to \infty$
corresponds to $X^9 \to -\infty$.
This results in the condition $a \neq 0$
and $c=0$.
Next by using the translation along the 
world volume direction and target space, 
we can set $b = 0$ and $a+d=0$, respectively.
Then, we obtain the following BPS configuration:
\begin{equation}
t = {1 \over z}.
\end{equation}
In terms of the original coordinates
$(X^9, X^{11})$ and the polar coordinate
on the world volume, 
$(x^1,x^2) = (r \cos \theta, r \sin \theta)$,
it is expressed as
\begin{align}
& X^9 = - R  \log r \,, \quad \\
& X^{11} = - R \theta \qquad ({\rm mod}\, 2 \pi R).
\end{align}

The fluctuation analysis around this configuration 
is exactly the same as in the previous subsection.
By requiring that the possible fluctuation
should be no more singular than the original one,
in particular both at $|z| \to 0$ and at $|z| \to \infty$, 
we find that the only possible candidate 
is the fluctuation of the overall constant, 
namely $R \to R + \delta R$.
However, such fluctuation conflicts with the 
periodicity of the target space and hence 
it is forbidden.
This result agrees with the physical 
intuition explained at the beginning of 
this subsection.

From the viewpoint of the 10-dimensional string theory,
the winding M2-brane solution considered in this section
is interpreted as a fundamental string ending on a D2-brane.
We will study the corresponding system in section \ref{D2}.

\section{Winding solution and fluctuation around it in the ABJM model}
In the previous section we studied the
fluctuation around the BPS configuration
of a single membrane on the flat target spacetime.
In this section we perform a similar analysis 
on the $\bm C^4/\bm Z_k$ orbifold background
based on the ${\rm U}(1) \times {\rm U}(1)$ ABJM model.

\subsection{$\boldsymbol{{\rm U}(1)} \times \boldsymbol{{\rm U}(1)}$ ABJM model}
We follow the convention of \cite{Bandres:2008ry},
in which the action of the
${\rm U}(1) \times {\rm U}(1)$ ABJM model 
is given by
\begin{equation}
S_{\rm ABJM} =
{k \over 2 \pi}
\int d^3 x 
\Big(
- D^\mu Y^A D_\mu Y_A
+
i \bar \Psi_A \gamma^\mu D_\mu \Psi^A
+
{1 \over 2} \epsilon^{\mu \nu \lambda}
\big(
A^{(1)}_\mu \partial_\nu A^{(1)}_\lambda
-
A^{(2)}_\mu \partial_\nu A^{(2)}_\lambda
\big)
\Big).
\label{S-ABJM}
\end{equation}
Here $Y^A$ and $Y_A$ ($A=1 \sim 4$) are 
the complex scalars and their complex conjugates.
The Dirac fermions $\Psi^A$ and $\Psi_A$
are also the complex conjugates of each other.
The lower (upper) index $A$ are the ${\bf 4}$ ($\overline{\bf 4}$) 
representation of the ${\rm SU}(4)$ R-symmetry.
The two ${\rm U}(1)$ gauge fields $A^{(1)}_\mu$ and $A^{(2)}_\mu$ 
are the Chern-Simons gauge fields with level $k$ and $-k$, respectively.
The fields $Y_A$ and $\Psi^A$ are the bi-fundamental representation
of the ${\rm U}(1) \times {\rm U}(1)$ 
gauge group with charge $(+,-)$.
Then the covariant derivatives for the scalars are defined 
as
\begin{align}
D_\mu Y^A & 
= 
\partial_\mu Y^A - i (A^{(1)}_\mu - A^{(2)}_\mu) Y^A\,, 
\label{DY^A}\\
D_\mu Y_A & 
= 
\partial_\mu Y_A + i (A^{(1)}_\mu - A^{(2)}_\mu) Y_A\,.
\label{DY_A}
\end{align}
Since the scalars and fermions are coupled only to 
the specific combination of the gauge fields, 
it is sometimes convenient to introduce the notation
$A^\pm_\mu \equiv A^{(1)}_\mu \pm A^{(2)}_\mu$.
With this, the Chern-Simons term is written as 
\begin{equation}
{1 \over 2} \epsilon^{\mu \nu \rho} 
(
A^{(1)}_\mu \partial_\nu A^{(1)}_\rho 
- 
A^{(2)}_\mu \partial_\nu A^{(2)}_\rho
)
=
{1 \over 2}\epsilon^{\mu \nu \rho} A^+_\mu \partial_\nu A^-_\rho
+
(\textrm{total derivatives})\,.
\end{equation}

This model is proposed as the IR limit of the 
world volume theory on a single ${\rm M}2$-brane 
on the fixed point of the $\bm C^4/\bm Z_k$ orbifold.
The three directions of the target space
are supposed to be identified 
with the world volume coordinates, $x^0$, $x^1$ and $x^2$ 
by taking the static gauge.
Then the remaining eight target space coordinates 
are combined into four complex scalars as 
\begin{align}
&
Y^1 = \sqrt{\pi T_{{\rm M}2} \over k} 
(X^3 + i X^4), \quad 
Y^2 = \sqrt{\pi T_{{\rm M}2} \over k}
(X^5 + i X^6), \nonumber \\
&
Y^3 = \sqrt{\pi T_{{\rm M}2} \over k}
(X^7 + i X^8), \quad 
Y^4 = \sqrt{\pi T_{{\rm M}2} \over k}
(X^9 + i X^{11}),
\label{XtoY}
\end{align}
where the normalization is fixed by comparing quadratic part of
the Nambu-Goto action \eqref{NG} and the ABJM model \eqref{S-ABJM}.
The $\bm Z_k$ orbifold identification is defined as 
$Y^A \sim {\rm e}^{2 \pi i \over k} Y^A$.

\subsection{Winding BPS M2-brane}
Let us consider an M2-brane solution which 
is winding around the orbifolding direction.
As in the previous section, 
we set $Y^{1,2,3} = 0$.
Then the BPS conditions coming from 
the requirement that the SUSY transformation of 
the gaugino fields $\Psi_A$ should vanish become
\begin{align}
\delta \Psi_1 & =
\gamma^\mu (- i \varepsilon^2 + \varepsilon^5) D_\mu Y^4 = 0\,, \\
\delta \Psi_2 & =
\gamma^\mu ( \varepsilon^1 + i \varepsilon^3) D_\mu Y^4 = 0\,, \\
\delta \Psi_3 & = 
\gamma^\mu ( - i \varepsilon^4 - \varepsilon^6) D_\mu Y^4 = 0\,,
\end{align}
where $\varepsilon^1$, \ldots, $\varepsilon^6$ are 
all two component Majorana spinors 
in three dimensions.
The condition coming from $\delta \Psi_4 = 0$ is
satisfied trivially.
Since these conditions can be summarized as 
$\gamma^\mu D_\mu Y^4 \varepsilon = 0$ with some 
spinor $\varepsilon$, the necessary condition is
given by\footnote{We have used $\gamma^0=i\sigma^2$,
$\gamma^1=\sigma^1$, $\gamma^2=\sigma^3$, where $\sigma^a$ 
are the Pauli matrices.}
\begin{align}
\det( \gamma^\mu D_\mu Y^4)
& = 
\det 
\left(
\begin{array}{cc}
D_2 Y^4 & D_0 Y^4 + D_1 Y^4 \\
- D_0 Y^4 + D_1 Y^4 & - D_2 Y^4
\end{array}
\right)\nonumber\\
& =
(D_0 Y^4)^2 - (D_1 Y^4)^2 - (D_2 Y^4)^2\nonumber \\
& = 0\,.
\end{align}
It is easy to check that this is also
the sufficient condition for the half-BPS configuration.
We further assume the static solution 
$\partial_0 Y^4 = 0$ and also 
$A^-_\mu=0$. The latter assumption is consistent
with the equation of motion.
Then the BPS condition is simplified as \cite{Sasaki:2009ij}
\begin{equation}
(\partial_1 + i \partial_2) Y^4 = 0, \quad
\textrm{or} \quad 
(\partial_1 - i \partial_2) Y^4 = 0,
\label{BPS-condition}
\end{equation}
namely the configuration $Y^4 = Y^4 (z, \bar z)$
is a half-BPS solution if it is either holomorphic or anti-holomorphic. 
In the rest of the paper, we consider the holomorphic case.

We specify the holomorphic function in a similar manner
as in the previous section.
A general form of the holomorphic function $Y^4 = Y^4(z)$ 
which is consistent with the orbifolding identification 
$Y^4 \sim {\rm e}^{2 \pi i \over k} Y^4$ 
can be studied by introducing the coordinate 
$t = (Y^4)^k$ 
and impose the condition $f(t,z)=0$.
First we consider a configuration which is
winding once around the orbifold direction.
By assuming also a unique zero of $f(t,z)$ 
for a fixed value of $z$, the form of $f(t,z)$
is restricted to
$f(t,z) = a t z + b t + c z + d$.
The coefficient $a$ is again not equal to zero
for a configuration which is reduced to a 
flat membrane in the $(X^1,X^2)$-plane 
in the limit $|z| \to \infty$.
The translational symmetry 
along the world volume direction is 
used to set $b=0$.
Since the solution with winding number one
satisfies $Y^4({\rm e}^{2 \pi i}z) 
= {\rm e}^{\pm {2 \pi i  \over k}} Y^4(z)$,
we impose $c=0$.
Then finally we end up with $f(t,z) = a tz + d = 0$, namely
$Y^4 \propto z^{ - {1 \over k}}$.
Now it is easy to generalize the configuration
to an M2-brane with winding number $n$ as
\begin{equation}
Y^4 = {\alpha \over z^{n \over k}}\,. \label{M2-spike}
\end{equation}
This configuration satisfies the winding property 
$Y^4({\rm e}^{2 \pi i} z) = {\rm e}^{-{2 \pi i n \over k}} Y^4(z)$.
The overall constant $\alpha$ can be taken 
to be real by using the phase of the $Y^4$ coordinate,
namely the rotation in the ($X^9$, $X^{11}$)-plane.
The equation of motion coming from the variation of \eqref{S-ABJM}
with respect to $A^-_\mu$ fixes the non-dynamical 
field $A^+_\mu$ in terms of the solution \eqref{M2-spike} as
\begin{equation}
A^+_0 = 2 Y^4 Y_4 + c, \quad A^+_1 = A^+_2 = 0.
\label{M2-sol-A+}
\end{equation}
The integration constant $c$ is introduced for the 
later convenience.

For large $k$, the orbifolding identification
effectively compactifies the $X^{11}$-direction 
and the above solution becomes much like a spike configuration 
extended in the $X^9$-direction. 
In fact, as we will see in the next section,
the winding solution is reduced to the 
D2-brane with a spike corresponding to the fundamental string. 

It is easy to check that the spike-like configuration \eqref{M2-spike} has 
the membrane tension $T_{{\rm M}2}$, 
which trivially follows from the normalization in \eqref{XtoY}. 
Thus \eqref{M2-spike} can be regarded as a deformation of the membrane itself. 
This is quite different from the situation in the spike solution on a D-brane 
discussed in \cite{Callan:1997kz}, where it is regarded not as the D-brane itself 
but as a fundamental string. Namely, our solution becomes massless when its thickness 
is zero, while the spike solution given in \cite{Callan:1997kz} still has a finite string tension 
even if it is sufficiently thin. We note here that in the latter 
the nontrivial gauge field configuration plays an essential role. 
We also notice that our normalization \eqref{XtoY} coincides with 
that adopted in \cite{Baek:2008ws} because we fix it essentially in the same way. 
What is important here is that under this normalization 
the one-loop effective action in the ABJM model also exactly agrees with the result 
of the supergravity computation up to the first nontrivial order of $v^4$ terms 
as shown in \cite{Baek:2008ws}.

\subsection{Fluctuations around BPS configuration}
We study fluctuations around 
the winding M2-brane configuration \eqref{M2-spike}.
As in the case of the Nambu-Goto action, 
we concentrate on the fluctuation 
of the field $Y^4$ with taking $Y^{1,2,3}=0$, while in the present case 
there exist also the fluctuations of the gauge fields
\begin{align}
& Y^1 = Y^2 = Y^3 = 0, \label{Y-fluc=0}\\
& Y^4 = Y_{\rm cl} + Y, 
\quad Y_4 = Y_{\rm cl}^\dag + Y^\dag, \label{Y-fluc} \\
& A_\mu^- = 0 + a_\mu^-, 
\quad A_\mu^+ = A_{{\rm cl},\mu}^+ + a_\mu^+. \label{A-fluc}
\end{align}
By inserting \eqref{Y-fluc=0}-- \eqref{A-fluc}
into \eqref{S-ABJM}, we obtain the quadratic part of the action 
for the fluctuations as 
\begin{align}
{\cal L} & = {k \over 2 \pi}
\bigg\{
- \partial^\mu Y \partial_\mu Y^\dag 
+ i 
\Big( 
Y \partial^\mu Y_{\rm cl}^\dag - Y^\dag \partial^\mu Y_{\rm cl}
+
Y_{\rm cl} \partial^\mu Y^\dag - Y_{\rm cl}^\dag \partial^\mu Y
\Big) a_\mu^- 
\notag \\
& \hspace{2cm}
- Y_{\rm cl}^\dag Y_{\rm cl} a^{-\mu} a^-_\mu 
+ {1 \over 2} \epsilon^{\mu \nu \rho} a^+_\mu \partial_\nu a^-_\rho
\bigg\} \,,
\label{L2}
\end{align}
where this action is valid when $Y$ is small compared to $Y_{\rm cl}$ 
or $k$ is large. The latter condition accords with the novel Higgs mechanism 
discussed in subsection \ref{section:NHM}. 
The conjugate momenta are 
\begin{align}
& P_Y = {\partial {\cal L} \over \partial \dot Y}
= {k \over 2 \pi}
\big( \dot Y^\dag - i Y_{\rm cl}^\dag a^{-0} \big) \,, 
\label{PY}\\
& P_{Y^\dag} = {\partial {\cal L} \over \partial \dot Y^\dag}
= {k \over 2 \pi} 
\big( \dot Y + i Y_{\rm cl} a^{-0} \big) \,, 
\label{PYd}\\
& P_{a^-}^\mu = {\partial {\cal L} \over \partial \dot a_\mu^-}
= {k \over 2 \pi} {1 \over 2} \epsilon^{0 \mu \nu} a_\nu^+ \,, 
\label{PA-}
\end{align}
and the canonical Hamiltonian for the fluctuations
is given by
\begin{align}
{\cal H} 
& = 
P_Y \dot Y 
+ P_{Y^\dag} \dot Y^\dag 
+ P_{a^-}^\mu \dot a_\mu^- - {\cal L} \notag \\
& = {k \over 2 \pi}
\bigg\{
\Big({2 \pi \over k} \Big)^2
P_{Y^\dag} P_Y
-
i
\Big(
Y \partial^\mu Y_{\rm cl}^\dag
-
Y^\dag \partial^\mu Y_{\rm cl}
+
Y_{\rm cl} \partial^\mu Y^\dag
-
Y_{\rm cl}^\dag \partial^\mu Y
\Big) a^-_\mu \notag \\
& \qquad 
+ Y_{\rm cl}^\dag Y_{\rm cl}
\big(
- 2 (a^{-0})^2 + (a^-_i)^2
\big)
+
\partial_i Y^\dag \partial_i Y
- 
{1 \over 2 }\epsilon^{\nu i \mu}
a^+_\nu \partial_i a^-_\mu
\bigg\}\,.
\end{align}

Now we want to find a zero mode by rewriting 
the Hamiltonian in terms of sum of squares. 
For this purpose we first note that 
if the gauge fields $a_\mu^\pm$ are integrated out in \eqref{L2} 
separately, they amount to imposing conditions 
\begin{align}
&
0 
=
i 
\Big( 
Y \partial^\mu Y_{\rm cl}^\dag
-
Y^\dag \partial^\mu Y_{\rm cl}
+
Y_{\rm cl} \partial^\mu Y^\dag
-
Y_{\rm cl}^\dag \partial^\mu Y
\Big)
- 2 Y_{\rm cl}^\dag Y_{\rm cl} a^{-\mu}
+ {1 \over 2} \epsilon^{\mu\nu\lambda} \partial_\nu a_\lambda^+, 
\label{ABJM-cons-1}\\
&
0 
= -{1 \over 2} \epsilon^{\mu \nu \lambda} \partial_\nu a_\lambda^-.
\label{ABJM-cons-2}
\end{align}
This suggests that it would be sufficient to look for a zero mode 
under these conditions. 
Then omitting total derivative terms, 
we have
\begin{align}
{\cal H}
& = 
{k \over 2 \pi}
\bigg\{
\Big( { 2 \pi \over k} \Big)^2 |P_Y|^2
- i 
\big(
Y \partial_i Y_{\rm cl}^\dag
-
Y^\dag \partial_i Y_{\rm cl}
+
Y_{\rm cl}  \partial_i Y^\dag
-
Y_{\rm cl}^\dag \partial_i Y
\big) a_i^- 
+
|Y_{\rm cl}|^2 (a_i^-)^2
+ 
|\partial_i Y|^2
\bigg\}.
\end{align}
We can bring it to a sum of square terms
by using that $Y_{\rm cl}$ is the holomorphic 
function of $z = x^1 + i x^2$.
Indeed the problematic second term 
can be rewritten as
\begin{equation}
- i 
(
Y \partial_i Y_{\rm cl}^\dag
-
Y^\dag \partial_i Y_{\rm cl}
+
Y_{\rm cl} \partial_i Y^\dag
-
Y_{\rm cl}^\dag \partial_i Y
) a^-_i
=
4i 
\{
(\partial_{\bar z} Y)Y_{\rm cl}^\dag a_z^-
-
(\partial_z Y^\dag) Y_{\rm cl} a_{\bar z}^-
\}
+(\textrm{total derivative})\,.
\end{equation}
Here, in addition to the holomorphy
$\partial_{\bar z} Y_{\rm cl} = 0$,
we have also utilized the constraint
$\partial_1 a^-_2 - \partial_2 a^-_1 = 0$.
Omitting the total derivative terms, the Hamiltonian
becomes the following form
\begin{equation}
{\cal H} 
=
{2 \pi \over k}
P_{Y^\dag} P_Y
+
{2 k \over \pi}
(\partial_z Y^\dag + i a_z^- Y_{\rm cl}^\dag)
(\partial_{\bar z} Y - i a_{\bar z}^- Y_{\rm cl})\,.
\label{H-LABJM}
\end{equation}
This clearly shows that the Hamiltonian
is semi-positive definite for the BPS configuration $\partial_{\bar z} Y_{\rm cl} =0$.
The flat direction can be specified by
\begin{equation}
\partial_{\bar z} Y - i a^-_{\bar z} Y_{\rm cl} = 0, \quad
P_{Y^\dag} = {k \over 2 \pi} (\dot Y - i Y_{\rm cl} a^-_0) = 0,
\label{flatdirection}
\end{equation}
with the conditions \eqref{ABJM-cons-1} and \eqref{ABJM-cons-2}.
Now it is clear that the static holomorphic function $Y=Y(z)$ 
with vanishing gauge fields $a^-_\mu = 0$ 
satisfies these equations.
The conditions \eqref{ABJM-cons-2}
are trivially satisfied and by solving
\eqref{ABJM-cons-1}, $a^+_j$ is determined. 
The solution of  $Y$ in \eqref{flatdirection} under \eqref{ABJM-cons-2} 
is shown to satisfy the equation of motion. 
For large $k$ and fixed $n$, the fluctuation $Y(z)$ should be 
no more singular than $Y_{\rm cl}(z)$, in particular both at the origin 
and at the infinity.  A generic form of such holomorphic functions 
which is consistent with the orbifolding condition 
$Y^4( {\rm e}^{2 \pi i} z) 
= {\rm e}^{-{2 \pi n i \over k}} Y^4(z)$
is given by $Y(z) = \Sigma_m c_m z^{-{n \over k}} z^m$. 
As a result we can single out 
a possible fluctuation as $Y(z) = z^{-{n \over k}}$.
This means the fluctuation of the overall factor
is again the possible marginal deformation,
$Y^4 = (\alpha + \delta \alpha) z^{-{n \over k}}$.

Remembering that the ABJM model is the conformal field theory,
the origin of the flat direction is clear.
It originates from the scale invariance 
which is now broken by the classical solution $Y_{\rm cl}(z)$. 
We may expect the existence of such a flat direction is rather robust. 
It is also natural that this fluctuation makes the configuration thinner as it does. 
This shows a contrast with the case of the zero modes originating 
from the translation and rotation where they do not deform the classical configuration.

\subsection{Nonlinear ABJM}
\label{nonlinearABJM}
Since our solution is singular at $|z|=0$, the argument based on the ABJM model 
might be modified if we take account 
of the nonlinear contributions.
In fact, a nonlinear extension of the 
ABJM model is proposed in \cite{Sasaki:2009ij}.
We address effect of the nonlinearity
on the analysis of the previous section.

The bosonic part of the nonlinear action
proposed in \cite{Sasaki:2009ij} is given by
\begin{equation}
S_{\textrm{ NL-ABJM}} = S_{\rm NG} + S_{\rm CS}.
\end{equation}
Here $S_{\rm NG}$ is given by
\begin{equation}
S_{\rm NG} = 
- 
T_{\rm M2} 
\int d^3 x
\sqrt{ 
- \det 
\Big(
\eta_{\mu \nu} 
+ 
{k \over 2 \pi T_{\rm M2}}
\big(
D_\mu Y^A D_\nu Y_A
+
D_\nu Y^A D_\mu Y_A
\big)
\Big)
}\,,
\label{SNG}
\end{equation}
where the covariant derivatives in \eqref{SNG} 
are the same as those in the linear ABJM model.
The second term $S_{\rm CS}$ is the same
Chern-Simons term as in \eqref{S-ABJM}.
If we set $Y^{1,2,3} = 0$, $S_{\rm NG}$ is simplified as
\begin{equation}
S_{\rm NG} 
= - T_{\rm M2} \int d^3 x
\sqrt{
1 
+ 
{k \over  \pi T_{\rm M2}}
|D_\mu Y^4|^2
-
\Big( 
{k \over 2 \pi T_{\rm M2}}
\Big)^2
\big(
(D_\mu Y^4)^2
(D_\nu Y_4)^2 
-
|D_\mu Y^4|^4
\big)
}\,.
\end{equation}
By considering the fluctuation as \eqref{Y-fluc}
and \eqref{A-fluc} with holomorphic $Y_{\rm cl}$, 
the quadratic action for the fluctuation is given by
\begin{align}
{\cal L} = 
&- {k \over 2 \pi}
\Bigg\{
|\partial_\mu Y|^2
+ | Y_{\rm cl}|^2 (a^-_\mu)^2
- i 
\big(
Y \partial^\mu Y_{\rm cl}^\dag 
- Y_{\rm cl}^\dag \partial^\mu Y
-({\rm c.c.})
\big) a^-_\mu 
- { 1 \over 2 }\epsilon^{\mu \nu \rho}
a^+_\mu \partial_\nu a^-_\rho
\notag \\
& \qquad 
-{2 k \over 2 \pi T_{\rm M2} + k |\partial_j Y_{\rm cl}|^2}
\big|
\partial_j Y_{\rm cl}
( \partial_j Y - i a_j^- Y_{\rm cl} )
\big|^2
\Bigg\} \,.
\label{S2-nonlinear}
\end{align}
Repeating a similar analysis from \eqref{L2} to \eqref{H-LABJM},  
we obtain the following Hamiltonian:
\begin{equation}
{\cal H}
=
{2 \pi \over k }P_Y P_{Y^\dag}
+
{k \over 2 \pi}
{4 \pi T_{\rm M2} \over \pi T_{\rm M2} + k |\partial_z Y_{\rm cl}|^2}
\big(
\partial_z Y^\dag + i Y_{\rm cl}^\dag a_z^-
\big)
\big(
\partial_{\bar z} Y - i Y_{\rm cl} a_{\bar z}^-
\big) \,.
\end{equation}
It is clear that the zero mode
analysis is not affected by the 
nonlinear extension of the ABJM model.

\section{Fundamental string attached to D2-brane}\label{D2}
  
In the previous section, we studied the zero mode 
around a winding M2-brane in the ABJM model.
In this section we examine that 
around a fundamental string for comparison.
We consider a bunch of fundamental strings ending on a D2-brane 
which is represented by a spike solution of the 
Dirac-Born-Infeld (DBI) action \cite{Callan:1997kz}. 
In subsection \ref{section:NHM} we will show that 
this spike D2-brane is actually related to 
the winding M2-brane \eqref{M2-spike}
through the novel Higgs mechanism.
Here we see that the winding number of the M2-brane 
around the orbifold corresponds to the number 
of the fundamental strings. 
This observation plays a crucial role 
in the later discussion on the quantum 
fluctuation around the spike solution.
In fact, we find a sharp difference between the 
zero mode analysis for an M2-brane and a fundamental string.
Notice that the fluctuation analysis in the present section 
should have implications on the stability 
of not a D2-brane but a fundamental string.

\subsection{Spike solution}\label{Spike solution}
We begin with a review of the spike solution derived in 
\cite{Callan:1997kz} in the D2-brane case.
This spike represents a bunch of fundamental 
strings ending on the D2-brane and it should
satisfy the charge quantization condition
corresponding to the string charge, 
namely the number of the strings.

The D2-brane world volume theory 
is given by the ${\rm U}(1)$ DBI action
\begin{align}
S_\text{D2}
& = - T_{\rm D2}\int d^3 x 
\sqrt{-\det (g_{\mu \nu}+ 2 \pi \alpha' F_{\mu \nu})} \\
& = - T_{\rm D2} \int d^3 x
\sqrt{ - \det (\eta_{\mu \nu} 
+ \partial_\mu X^i \partial_\nu X^i
+ 2 \pi \alpha' F_{\mu \nu})} \,,
\label{D2 action}
\end{align}
where the tension of the D2-brane is defined by
$T_{\rm D2} =(2 \pi)^{-2} {\alpha'}^{-{3 \over 2}} g_s^{-1}$
and $g_s$ is the string coupling constant. 
$g_{\mu \nu}$ and $F_{\mu \nu}$ 
are the induced metric and the ${\rm U}(1)$ 
gauge field strength, respectively.
$\mu,\nu=0,1,2$ are world volume indices.
In the second line, we took the static gauge 
and $X^i$ ($i=3,\ldots,9$) are the transverse target coordinates.

The spike solution in \cite{Callan:1997kz} has a nontrivial 
configuration only for one of the scalars, say $X^9$, and 
also the gauge potential $A_0$. 
Since the solution corresponds to D2-F1 system
which preserves the eight supersymmetry,
its derivation is based on the study of the BPS condition.
As we explain in the appendix \ref{SUSY-D2}, 
if we assume the ${\rm S}^1$ symmetric static ansatz, 
$X^9=X^9(r)$ and $F_{0r} =F_{0r}(r)$, then the configuration 
preserves half of the supersymmetry provided it satisfies 
the condition $(\partial_r X^9)^2 = (2 \pi \alpha' F_{0r})^2$.
The specific solution which indeed represents 
a fundamental string ending on a D2-brane is given by 
\begin{align}
X^9 = 2 \pi \alpha' A_0 = c\log\frac{r}{\ell}, \label{gen solution}
\end{align}
where $c$ and $\ell$ are dimensionful constants. 

The overall constant $c$ needs to be tuned so that 
\eqref{gen solution} represents fundamental strings. 
Moreover, $c$ is quantized corresponding to 
the string charge.
In order to find the correct quantization,
we study how the spike solution is coupled 
to the NS-NS $B$ field. 
In the case of the $n$ fundamental strings extended 
into $X^9$ direction, the string world sheet is coupled 
to the $B$ field as
\begin{equation}
n T_{\rm F} \int dX^0 dX^9 B_{09},
\end{equation}
where $T_{\rm F}=1/(2\pi\alpha')$ and we identified $X^0$ and $X^9$ 
with the world volume coordinates.
Next we consider the case of the above spike D2-brane.
Let us introduce the notation 
\begin{equation}
S_{\rm D2}
= - T_{\rm D2} \int dt dr d \theta r
\sqrt{- \det(g_{\mu \nu} +B_{\mu\nu}+2 \pi \alpha' F_{\mu \nu})}
=
\int dX^0 dX^9 {\cal L},
\end{equation}
where $B_{\mu\nu}$ is the $B$ field induced on the world volume. 
We assumed ${\rm S}^1$ symmetric ansatz $dX^9 = d r \partial_r X^9$.
Then in order for the spike solution to 
represent the $n$ fundamental strings, 
the differential of $\cal L$ with respect to 
$B_{09}$ should be equal to the string charge\footnote{
The resulting flux quantization condition is 
the same as the one given in \cite{Callan:1997kz}.}
\begin{equation}
n 
= 
\left.T_{\rm F}^{-1}{\partial {\cal L} \over \partial B_{09}}\right|_{B=0}
=
\left.\partial_r X^9(r){\partial {\cal L} \over \partial F_{0r}}\right|_{B=0}
= (2 \pi \alpha')^2 T_{\rm D2} \int_{\rm S^1} F_{0r},
\end{equation}
where in the last equality we used the BPS condition 
$(\partial_r X^9(r))^2 = (2 \pi \alpha' F_{0r}(r))^2$. 
For the spike solution \eqref{gen solution},
this means the following quantization of the 
factor $c$\footnote
{Here the sign is not important because it depends 
on the orientation of the string and the D2-brane.} 
\begin{equation}
n = - (2 \pi)^2 \alpha' T_{\rm D2} c 
\quad 
\to  
\quad 
c = - n g_s \ell_s .
\end{equation}
Consequently, the half BPS solution of the DBI action representing 
$n$ fundamental strings is
\begin{align}
X^9 = 2 \pi \alpha' A_0 
= - n g_s \ell_s  \log\frac{r}{\ell}.
\label{spike}
\end{align}
Since the combination $R = g_s \ell_s$ is the radius of the 
M-theory circle the solution corresponds to 
the M2-brane solution winding around the M-theory circle
which is discussed in \ref{Compact target space}.

Finally, we calculate the energy of the spike 
to confirm that it can be identified 
with a fundamental string. From the DBI action 
\eqref{D2 action} its energy\footnote
{Note that in order to obtain the energy of the spike, 
we have to subtract the energy of the D2-brane itself from 
the  Hamiltonian evaluated with the spike solution. } is
\begin{align}
E = 
T_{\rm D2}
\int^{\Lambda}_{\delta}dr d\theta r 
\Big({n \ell_s g_s \over r}\Big)^2 
= nT_{\rm F}\Big[X^9(r=\delta)-X^9(r=\Lambda)\Big],
\label{spike energy}
\end{align}
where we have introduced cutoffs to regularize divergences 
both at the origin and infinity of $r$.
The energy of the spike \eqref{spike energy} tells us that $n$ is in fact interpreted 
as the number of fundamental strings, 
because the energy of the spike with $n=1$ 
is exactly the tension of a fundamental string 
times its length.

\subsection{Novel Higgs mechanism}\label{section:NHM}
In this subsection we clarify how the spike solution with 
string (electric) charge $n$ discussed in the previous 
subsection is related to the half-BPS solution in the ABJM model 
\eqref{M2-spike}, \eqref{M2-sol-A+} 
via the novel Higgs mechanism.  

\subsubsection{A brief review} 
Let us begin with a brief review of 
the novel Higgs mechanism in the case of 
${\rm U}(1)\times {\rm U}(1)$ ABJM model.
For simplicity, we consider only 
the bosonic part of the ABJM model
\begin{equation}
S = {k \over 2 \pi}
\int d^3 x \Big( - D^\mu Y^A D_\mu Y_A + 
{1 \over 2} 
\epsilon^{\mu \nu \rho} 
A^+_\mu \partial_\nu A^-_\rho \Big).
\end{equation}
We assume a vacuum expectation value $v$ 
for the real component of
$\sqrt{k/(2\pi)}Y^4$ and rename the fields as
\begin{align}
&\sqrt{k \over 2 \pi} Y^a 
= 
\sqrt{{T_{{\rm M}2} \over 2}}
( \widetilde X^{2a+1} + i \widetilde X^{2a+2})  
\quad (a=1,2,3), \\
& 
\sqrt{k \over 2 \pi} Y^4 = v + 
\sqrt{{T_{{\rm M}2} \over 2}}
(\widetilde X^9 + i \widetilde X^{11}), 
\label{Higgs-X} \\
& A_\mu^- = {B_\mu \over v},\quad 
A_\mu^+ = 2 A_\mu. \label{Higgs-A}
\end{align}
Here $\widetilde X^i$ is related to the target space coordinate $X^i$ 
by the shift as $ X^i =  \sqrt{2 / T_{{\rm M}2}} v \delta^{i 9} +
\widetilde X^i $.
Then by taking large $k$ and $v$ limit with keeping 
${v \over k}$ finite,\footnote{
As mentioned just before the equation 
\eqref{Mtheory2Stringtheory}, the scale of the ratio ${v \over k}$ 
is given by
${v \over k} = R \sqrt{T_{\rm M2} \over 2} \sim ({g_s \over \ell_s})^{1
\over 2}$. 
Hence the condition that ${v \over k}$ is kept finite
means that we take the scale $g_s \over \ell_s$ to be finite
and that in this unit all dimensionful quantities are described.
\label{footnote-unit}
}
the action becomes
\begin{align}
S  = \int d^3 x
& \bigg\{
-{T_{\rm M2} \over 2}
(
\partial^\mu \widetilde X^i 
\partial_\mu \widetilde X^i + 
\partial^\mu \widetilde X^{11} 
\partial_\mu \widetilde X^{11})
-B^\mu B_\mu \notag \\
& \hspace{1cm}
+ \sqrt{2 T_{\rm M2}} \partial^\mu \widetilde X^{11} B_\mu
+ {k \over 2 \pi v} \epsilon^{\mu \nu \rho}
A_\mu \partial_\nu B_\rho
\bigg\}, 
\label{ABJM-2-D2}
\end{align}
where the summation over $i$ is taken for $i=3 \sim 9$.

Next, by using the equation of motion for $ B_\mu$
\begin{equation}
B_\mu = 
\sqrt{T_{\rm M2}\over 2}
\partial_\mu \widetilde X^{11} 
+ 
{k \over 4 \pi v}
\epsilon_{\mu \nu \rho} \partial^\nu A^\rho\,,
\end{equation}
and neglecting the boundary terms, the action becomes
\begin{equation}
S = -
\int d^3 x
\Big(
{T_{\rm M2} \over 2}
\partial^\mu \widetilde X^i 
\partial_\mu \widetilde X^i
+
{1 \over 2} 
\Big({k \over 4 \pi v} \Big)^2
F_{\mu \nu} F^{\mu \nu}
\Big).
\label{ABJM-NHM-S1}
\end{equation}
Now recall that around $X^9 \sim \sqrt{2/T_{\rm M2}} v$,
the radius of the M-theory circle is 
given by $R=\sqrt{2 \over T_{\rm M2}}{v \over k}$ 
\cite{Mukhi:2008ux,Ho:2008ei,Honma:2008jd,Pang:2008hw}. 
Then by using the relation 
\begin{equation}
g_s = (R / \ell_{\rm P})^{3/2},\quad
\alpha' = R^{-1} \ell_{\rm P}^3,
\label{Mtheory2Stringtheory}
\end{equation}
and the definition of the membrane tension
$T_{\rm M2} = (2\pi)^{-2} \ell_{\rm P}^{-3}$, 
\eqref{ABJM-NHM-S1} becomes
\begin{equation}
S = - {\ell_s \over g_s}\int d^3x 
\Big(
{1 \over 2} \partial^\mu \Phi^i \partial_\mu \Phi^i
+
{1 \over 4} F_{\mu \nu} F^{\mu \nu}
\Big), \label{YM action}
\end{equation}
where the scalar field $\Phi^i$ is defined in the standard manner by
\begin{equation}
\Phi^i = {1 \over 2 \pi \alpha'} \widetilde X^i,
\end{equation}
and the overall factor $\ell_s / g_s$ should be
identified with the three-dimensional Yang-Mills 
coupling constant as $\ell_s /g_s = 1/g_{\rm YM}^2$.

\subsubsection{Novel Higgs mechanism for the winding solution}

Next we consider the novel Higgs mechanism 
for the M2-brane solution 
with winding number $n$, \eqref{M2-spike} and 
\eqref{M2-sol-A+}, 
\begin{align}
&Y^A=0\,~(A=1\sim 3), ~~~A_\mu^-=0, \nonumber \\
&Y^4=Y_{\text{cl}}=\alpha z^{- {n\over k}}, \nonumber \\
&A_0^+=2Y_\text{cl}Y_\text{cl}^\dagger+c,~~~A_1^+=A_2^+=0.
\label{M2-spike2}
\end{align}
Namely, we interpret it in terms of 
the D2-brane world volume fields
$\Phi^i$ and $A_\mu$ in the limit 
$k \to \infty$.
In the limit, $\sqrt{k /(2 \pi)}Y_{\rm cl}$ is expanded as 
\begin{equation}
\sqrt{k \over 2\pi}Y_{\rm cl} (z) 
=
\sqrt{k \over 2 \pi} \alpha
\Big(
1
- {n \over k}\log r
- i {n \over k} \theta
+ {\cal O}(k^{-2})
\Big).
\label{Y_cl-large-k}
\end{equation}
The novel Higgs mechanism  
tells us that 
if the real part of $\sqrt{k/(2 \pi)} Y^4$ develops
the large constant expectation value $v$ ($\sim k$),
the remaining ${\cal O}(1)$ part of it 
can be identified as the scalar field $\Phi^9$ 
on the D2-brane world volume. 
Therefore we consider the large $\alpha$ and 
identify it as 
\begin{equation}
\sqrt{k \over 2 \pi} \alpha = v \to \infty.
\end{equation}
Then the remaining part of \eqref{Y_cl-large-k} 
can be regarded as the classical configuration for
the scalar field $\Phi^9$ and the compact direction as 
\begin{equation}
\widetilde X^9 \rightarrow - { ng_s\ell_s}\log r \,, \quad
\widetilde X^{11} \rightarrow - {ng_s\ell_s}\theta\,, 
\label{M2-D2-X}
\end{equation}
where we used $\frac{v}{k}=\sqrt{\frac{T_{\rm M2}}{2}}R$. 
Similarly the D2-brane gauge field 
$A_\mu$ can be deduced from \eqref{Higgs-A} and
\eqref{M2-spike2} as
\begin{align}
A_0 
& = Y_\text{cl}Y^\dagger_\text{cl}+\frac12 c
= \alpha^2 z^{-{n \over k}} \bar z^{-{n \over k}}
+ {1 \over 2} c \label{A=YY+c/2} \\
& = {2 \pi v^2 \over k} 
\Big(
1 - 2 {n \over k} \log r + {\cal O}(k^{-2}) 
\Big)
+ {1 \over 2} c \,.
\end{align}
This means 
\begin{equation}
A_0 \rightarrow 
- {g_s \over 2 \pi \ell_s} n \log r,
\label{M2-D2-A}
\end{equation}
where we have chosen $c = - 4 \pi v^2 /k$ for simplicity, 
because a constant of the gauge field does not play any role.
Since the configuration for the fields 
$Y^1$, $Y^2$ and $Y^3$ are trivial,
the remaining scalar fields are zeros
$\Phi^i=0$ ($i=3 \sim 8$).
Recalling the computation \eqref{spike energy}, 
the resulting spike D2-brane solution obviously
reproduces the correct string tension.

From \eqref{M2-D2-X} and \eqref{M2-D2-A},
we see that our winding solution \eqref{M2-spike2} 
is correctly reduced to the spike solution on a D2-brane given in \eqref{spike} 
as long as configurations around $X^9=\sqrt{2\over T_{\rm M2}}v$ are concerned. 
In fact, they coincide 
even including the coefficients. 
It should be noticed that the winding number $n$ 
appears as the string charge after the novel Higgs mechanism. 
This is consistent with the M-theory interpretation 
of a fundamental string. 
Here we remember that 
the string charge appears as the overall factor of the solution, 
while in the case of the winding M2-brane solution \eqref{M2-spike2}, 
the winding number is encoded into the power. 
In the next subsection, we see that a possible zero mode 
of fluctuations around the spike on the D2-brane can 
shift slightly the overall factor of the original spike solution. 
However, such a zero mode is forbidden  
in contrast to the M2-brane case
because we have fixed the string charge 
as the winding number of the M2-brane. 
Therefore, although the spike solutions on the M2-brane and 
D2-brane are directly related by the novel Higgs mechanism 
as we have seen above, 
the difference in their dependence of the charge $n$ 
yields a clear difference in existence of a zero mode.

\subsection{The stability of a fundamental string}
\label{subsection:string-stability}

Let us consider fluctuations around the spike solution 
\eqref{spike} in \eqref{D2 action} 
and examine stability of a fundamental string attached 
to a D2-brane. Then we point out a difference between an M2-brane 
and a fundamental string.

Fluctuations we are interested in are 
\begin{equation}
{X^9 \over 2 \pi \alpha'}
={R \over 2 \pi \alpha'}\log\frac{r}{\ell} + y(t,r,\theta),\quad 
A_0 = {R \over 2 \pi \alpha'}\log\frac{r}{\ell} + \phi(t,r,\theta),\quad 
A_r(t,r,\theta),\quad A_{\theta}(t,r,\theta),
\label{D2fluctuations}
\end{equation}
where for later convenience, we take the polar coordinate system.
Substituting these fluctuations for the action (\ref{D2 action}) 
and keeping terms of second order with respect to them, 
the resulting quadratic action is
\begin{eqnarray}
S_2&=&\frac{\ell_s}{g_s}\int dt dr d\theta r \Bigl[\frac{1}{2}\dot{y}^2 - \frac{1}{2}y'^2 - \frac{1}{2r^2}(\partial_{\theta}y)^2
- \frac{1}{2r^2}(A'_{\theta}-\partial_{\theta}A_r)^2 + \frac{1}{2r^2}(\dot{A}_{\theta}-\partial_{\theta}\phi)^2
\nonumber\\
&& + \frac{1}{2}(\dot{A}_r-\phi^{\prime})^2
+ \frac{R^2}{2r^4}(\dot{A}_{\theta}+\partial_{\theta}y-\partial_{\theta}\phi)^2 + \frac{R^2}{2r^2}(\dot{A}_r+y'-\phi^{\prime})^2\Bigl].
\label{fluc L for D2}
\end{eqnarray}
When $R=0$, $S_2$ becomes the Yang-Mills action. 
It is easy to see that up to this order the other fluctuations 
than those in \eqref{D2fluctuations} decouples. 
We consider the Hamiltonian obtained from $S_2$ 
\begin{eqnarray}
H_2&=&\int dr d\theta \Bigl\{P_y \dot{y} + P_{A_r} \dot{A_r} + P_{A_{\theta}} \dot{A}_{\theta} - \cal{L}\Bigl\}\nonumber\\
&=& \int dr d\theta \Bigl\{\frac{g_s}{2\ell_s r}P_y^2 + \frac{\ell_s}{2g_sr}F_{r\theta}^2 
+\frac{\ell_s}{2g_sr}\Bigl(1+\frac{R^2}{r^2}\Bigl)^{-1}
\Big({g_s \over \ell_s}P_{A_r} + ry^{\prime}\Big)^2 \nonumber\\
&&+\frac{\ell_s r}{2g_s}\Bigl(1+\frac{R^2}{r^2}\Bigl)^{-1}
\Bigl({g_s \over \ell_s}P_{A_{\theta}} + \frac{1}{r}\partial_{\theta}y\Bigl)^2 
+P_{A_r}\phi^{\prime}+P_{A_{\theta}}\partial_{\theta}\phi - P_{A_{\theta}}\partial_{\theta}y - P_{A_r}y^{\prime}\Bigl\}, \notag \\
\label{D2 Hamiltonian}
\end{eqnarray}
where the conjugate momenta are
\begin{eqnarray}
P_{y}&=&\frac{\partial \cal{L}}{\partial \dot{y}}=\frac{\ell_s r}{g_s} \dot{y} \label{can y},\\
P_{A_r}&=&\frac{\partial \cal{L}}{\partial \dot{A}_r}
=\frac{\ell_s r}{g_s}\Bigl\{\Bigl(1+\frac{R^2}{r^2}\Bigl)(\dot{A}_r-\phi^{\prime})+\frac{R^2}{r^2}y^{\prime}\Bigl\}\label{can Ar},\\
P_{A_{\theta}}&=&\frac{\partial \cal{L}}{\partial \dot{A}_{\theta}}
=\frac{\ell_s r}{g_s}\frac{1}{r^2}\Bigl\{\Bigl(1+\frac{R^2}{r^2}\Bigl)(\dot{A}_{\theta}-\partial_{\theta}\phi)+\frac{R^2}{r^2}\partial_{\theta}y\Bigl\}.
\label{can Ath}
\end{eqnarray}
From variations of (\ref{D2 Hamiltonian}) with respect to $\phi$, the Gauss' law constraint reads
\begin{eqnarray}
0&=&\partial_r P_{A_r} +\partial_{\theta}P_{A_{\theta}}\nonumber\\
&=&  \frac{\ell_s}{g_s}\Bigl[\partial_r\Bigl\{r\Bigl(1+\frac{R^2}{r^2}\Bigl)(\dot{A}_r-\phi^{\prime})+\frac{R^2}{r}y^{\prime}\Bigl\}
+\partial_{\theta}\Bigl\{\frac{1}{r}\Bigl(1+\frac{R^2}{r^2}\Bigl)(\dot{A}_{\theta}-\partial_{\theta}\phi)+\frac{R^2}{r^3}\partial_{\theta}y\Bigl\}\Bigl] .
\label{D2 gauss}\nonumber\\
\end{eqnarray}
Using the partial integration and the Gauss' law constraint, 
the Hamiltonian becomes
\begin{eqnarray}
H_2&=& \int dr d\theta \Bigl\{\frac{g_s}{2 \ell_s r}P_y^2 
+ \frac{\ell_s}{2g_sr}F_{r\theta}^2 
+\frac{\ell_s}{2g_sr}\Bigl(1+\frac{R^2}{r^2}\Bigl)^{-1}
\Big({g_s \over \ell_s}P_{A_r} + ry^{\prime}\Big)^2 \nonumber\\
&&+\frac{\ell_s r}{2g_s}\Bigl(1+\frac{R^2}{r^2}\Bigl)^{-1}
\Bigl({g_s \over \ell_s}P_{A_{\theta}} 
+ \frac{1}{r}\partial_{\theta}y\Bigl)^2\Bigr\},
\label{D2 Hamiltonian2}
\end{eqnarray}
where contributions from boundary terms are ignored.
Thus the Hamiltonian is positive semi-definite 
up to boundary contributions. 
In appendix~\ref{app:bc}, we will consider a contribution 
from a boundary to the Hamiltonian.

Next, we consider the minimum of the Hamiltonian 
and find a zero mode. From (\ref{D2 Hamiltonian2}) 
$H_2$ has the minimum when
\begin{eqnarray}
P_y=0,~~F_{r\theta}=0,~~P_{A_r}=-\frac{\ell_s r}{g_s}y',~~P_{A_{\theta}}=-\frac{\ell_s}{g_s r} \partial_{\theta}y.
\label{D2zeromodecond}
\end{eqnarray}
Fluctuations that satisfy these conditions 
and the Gauss' law constraint are shown to satisfy the equation of motion 
and also are candidates for zero modes.
Let us first find a zero mode of the fluctuation $y$. 
Using $P_y=0$ and (\ref{can y}), $y$ is time-independent:
\begin{eqnarray}
y=y(r,\theta). 
\end{eqnarray}
From the last two equations in \eqref{D2zeromodecond} 
and $(\ref{D2 gauss})$,
we find that $y(r,\theta)$ should satisfy the Laplace equation: 
\begin{eqnarray}
r\partial_r(r\partial_r y)+\partial_{\theta}^2y=0,
\label{laplace}
\end{eqnarray}
whose solution is given by 
\begin{eqnarray}
y(r,\theta)=\sum^{\infty}_{m=-\infty\atop m\neq 0}(y_m r^m e^{im\theta} + y^*_m r^m e^{-im\theta}), \label{sol m}
\end{eqnarray}
where we have imposed the reality condition. 
In general fluctuations must be smaller 
than the classical solution \eqref{spike} 
both at the origin and infinity within our approximation 
of taking only the second order of fluctuations. 
Then all fluctuations in \eqref{sol m} are not allowed 
because they are more singular than \eqref{spike} 
either at the origin or infinity. 
However, if we relax \eqref{laplace} at the origin where 
the original classical solution itself is ill-defined, 
we may allow a solution of \eqref{laplace} 
corresponding to $m=0$ case in \eqref{sol m}
\begin{eqnarray}
y(r,\theta)=\alpha\log\frac{r}{\ell}, \label{sol m=0}
\end{eqnarray}
where $\alpha \ll R$ is a constant. From now on, we consider 
\eqref{sol m=0} as a candidate for the zero mode of $y$.

Let us find zero modes of the fluctuations $A_0,A_r,A_{\theta}$.
From $F_{r\theta}=0$, $A_r$ and $A_\theta$ can be rewritten as 
\begin{eqnarray}
A_{r}=\partial_r a(t,r,\theta),~~A_{\theta}=\partial_{\theta}a(t,r,\theta),
\end{eqnarray}
using a scalar field $a(t,r,\theta)$. 
Then $P_{A_r}=-\frac{\ell_s r}{g_s}\partial_r y$, (\ref{can Ar}) 
and $A_r=\partial_r a$ yield 
\begin{eqnarray}
\phi= y + \partial_t a -f(t,\theta),
\end{eqnarray}
where a function $f$ is independent of $r$. 
Furthermore, 
$P_{A_{\theta}}=-\frac{\ell_s}{g_s r}\partial_{\theta}y$ and 
(\ref{can Ath}) tell us that $f$ depends only on time: $f=f(t)$.
In summary we find that a candidate for the zero mode 
is given in general as 
\begin{align}
y=\alpha\log\frac{r}{\ell},~~~
\phi=y+ \partial_t a -f(t),~~~
A_r=\partial_r a,~~~
A_{\theta}=\partial_{\theta}a.
\end{align} 
By a gauge transformation, these equations can be simplified 
into 
\begin{eqnarray}
y=\phi=\alpha\log\frac{r}{\ell},~~A_r=0,~~A_{\theta}=0. 
\label{zero mode}
\end{eqnarray}
These imply that fluctuations giving the minimum 
of the Hamiltonian preserve the half-BPS condition, 
which is similar to the winding M2-brane case in the ABJM model. 
In fact if we plug \eqref{zero mode} into $S_2$, 
we find that it vanishes up to boundary contributions\footnote
{However, in the case of \eqref{zero mode} 
boundary terms have nontrivial values in general 
according to boundary conditions. See appendix \ref{app:bc}.}. 
Thus it may be possible that 
this mode would be a flat direction 
and indicate instability of the spike solution. 
However there is really an important difference 
between an M2-brane and a fundamental string. 
Namely, the spike on the D2-brane 
corresponds to a bunch of fundamental strings, 
where the number of them are related to the overall factor 
of the spike solution. 
Since \eqref{sol m=0} takes exactly the same form 
as the classical solution \eqref{spike}, it makes a slight change 
of the overall factor of the solution, or equivalently 
the number of strings. Hence such a mode cannot be allowed 
under the condition of fixed string charge. 
On the other hand, in the ABJM model the winding number 
is not related to the overall factor of the solution 
\eqref{M2-spike} and consequently the zero mode which 
takes the same form as the classical solution 
is allowed unlike the case of a fundamental string. 
We can, therefore, reasonably conclude 
that a fundamental string ending on a D2-brane is stable at least perturbatively, 
because a zero mode of fluctuations does not exist.

Finally, we comment on zero modes related to symmetries. 
the original DBI action \eqref{D2 action} has 
the translational symmetry on the world volume which the spike solution 
\eqref{spike} breaks. 
As a consequence, a zero mode associated with this breaking 
should exist.
It is actually included in the solution \eqref{sol m}, 
but it is forbidden from behaviour of divergence at the origin. 
Likewise, a zero mode associated with a broken symmetry would be 
in general more singular at the origin 
than the classical solution itself, because it would correspond 
to a deformation of the classical solution by the broken symmetry 
and therefore take a form of a derivative of the classical solution. 
So we cannot find an allowed zero mode related to a symmetry 
which the spike breaks. 
This situation is quite different from the winding M2-brane 
in that it has a zero mode related to 
the broken scale symmetry, 
which does not make the classical solution more singular. 

\subsection{Novel Higgs mechanism for zero mode}
In the analysis of the ABJM model, we found the zero mode
which deforms the original spike-like configuration, while
in subsection \ref{subsection:string-stability}
we found no such fluctuation around 
the spike D2-brane.
Since these two solutions are related through 
the novel Higgs mechanism, one may ask how 
the zero mode is mapped.
In order to clarify it, we consider the deformation 
$Y^4(z) = (\alpha + \delta \alpha) z^{-{n \over k}}$
and reexamine the novel Higgs mechanism.
We concentrate on the real $\delta \alpha$ since the imaginary
part corresponds to the rotation of the configuration
and does not deform it.

The expansion of $\sqrt{k/(2 \pi)} Y^4$ is now given by
\begin{align}
\sqrt{ k \over 2 \pi} Y^4(z) 
& =
v 
\Big(
1 + {\delta \alpha \over \alpha}
\Big)
\Big(
1 - {n \over k} \log r - i {n \over k} \theta + {\cal O}(k^{-2})
\Big).
\end{align}
Here $v$ is defined by $ v = \sqrt{k/(2\pi)} \alpha$ as before.
By subtracting the large expectation value $v$, 
the coordinates $\widetilde X^9$ and $\widetilde X^{11}$ 
are identified as 
\begin{align}
& 
\sqrt{T_{\rm M2} \over 2}
\widetilde X^9  
= 
- n { v \over k} \log r 
+ v {\delta \alpha \over \alpha } 
- {\delta \alpha \over \alpha } n {v \over k}\log r 
+ \cdots
, \label{wideX9}
\\ 
&
\sqrt{T_{\rm M2} \over 2}
\widetilde X^{11}  
= 
- n {v \over k} \theta 
- {\delta \alpha \over \alpha } n {v \over k} \theta 
+ \cdots
. \label{wideX11}
\end{align}
The first terms in these equations are 
the classical configurations given in \eqref{M2-D2-X}.
When we derived \eqref{ABJM-2-D2}, we assumed that 
the order of the fields are as 
$\sqrt{T_{\rm M2}} \widetilde X^i \sim {\cal O}(1)$.\footnote{
Recall this order counting is under the fixed
scale ${v \over k}$ as mentioned in footnote \ref{footnote-unit}.
}
Hence as long as $k$ and $v$ dependences are concerned, 
the leading fluctuation, namely the constant term
$ v {\delta \alpha \over \alpha}$ in \eqref{wideX9} 
should be of the same order, ${\cal O}(1)$.
Otherwise the fluctuation $\delta \alpha$ 
cannot be regarded as the fluctuation 
in the D2-brane theory.
This requirement is equivalent to the condition
${\delta \alpha \over \alpha} = {\cal O}(k^{-1})$
and then the remaining terms 
in \eqref{wideX9} and \eqref{wideX11} vanish in the large $k$ limit.
This clearly shows that the zero mode found in the 
ABJM model corresponds to the constant shift along 
the $\widetilde X^9$ direction in the D2-brane theory.
In terms of $y$ given in \eqref{D2fluctuations}, 
the fluctuation is expressed as $y = 2 \alpha \delta \alpha$.

In the similar manner, the corresponding fluctuation for 
the gauge field $A_0$ can be deduced from \eqref{A=YY+c/2}.
By taking account of the fluctuation for $Y^4$
as 
$Y^4 = Y_{\rm cl}+Y = (\alpha + \delta \alpha) z^{-{n\over k}}$
we have 
\begin{align}
A_0 
& = 
\big(Y_{\rm cl} + Y\big)
\big(Y_{\rm cl}^\dag + Y^\dag\big)
 - {2 \pi v^2 \over k}  \notag \\
& =
\Big( Y_{\rm cl} Y_{\rm cl}^\dag - {2 \pi v^2 \over k} \Big)
+Y_{\rm cl} Y^\dag + Y Y_{\rm cl}^\dag + Y Y^\dag\\
& \to 
- {g_s \over 2 \pi \ell_s} n \log r + 2 \alpha \delta \alpha.
\quad (k \to \infty) \label{NHM(A0+fluc)}
\end{align}
The first term in \eqref{NHM(A0+fluc)}
is the classical configuration \eqref{M2-D2-A} 
and the remaining constant shift $2 \alpha \delta \alpha$ 
is the fluctuation corresponding to $\phi$ 
defined in \eqref{D2fluctuations}.

One should notice that the above result dose not mean 
that the constant shift along the $\widetilde X^9$ direction 
in the D2-brane theory implies any instability contained in 
the D2-brane theory itself.
The novel Higgs mechanism connects the ABJM model 
and the D2-brane theory only locally around 
$\widetilde X^9 = \sqrt{2 \over T_{\rm M2}} v$.
Therefore the precise meaning of the above observation 
is that, around this region, the deformation of the M2-brane 
is described by the constant shift of the D2-brane. 
On the other hand, the constant shift in the D2-brane theory 
itself dose not deform the shape of the spike D2-brane, 
and hence there is no reason to expect that 
it implies the instability of the fundamental string.

As a final check, if we allow a small change 
of the winding number in the ABJM model as $Y^4(z)=\alpha z^{-{n+\delta
n \over k}}$, then under the novel Higgs mechanism, 
it is reduced to the change of the overall constant of 
the fields as 
$X^9 = 2 \pi \alpha' A_0 = - (n+\delta n) g_s \ell_s \log r$ 
in the D2-brane theory.
This means that such fluctuations are forbidden by essentially 
the same reason both in the ABJM model and the D2-brane theory.

\section{Conclusions and discussions}
In order to study the instability of a membrane and a string, 
we first constructed half-BPS solutions in the Nambu-Goto action 
of a supermembrane, the ${\rm U}(1)\times {\rm U}(1)$ ABJM model, 
and the DBI action of the D2-brane. Then we looked for zero modes 
which deform them under fixed charges, 
namely the winding number for a wrapped membrane, 
or the number of strings. In the case of a supermembrane, 
a zero mode is indeed found even if we fix its winding number. 
In particular, in the case of 
the ${\rm U}(1)\times {\rm U}(1)$ ABJM model with large $k$, 
the BPS solution becomes a configuration like the spike 
and the zero mode we found can make it thinner. This is 
in accordance with the physical picture proposed 
in \cite{deWit:1988ct}. On the other hand, the spike solution 
in the DBI action of the D2-brane does not allow a zero mode 
under a fixed string charge. 
The situation is similar to the Nambu-Goto action 
of a compactified supermembrane with the winding number fixed. 
They are consistent with the fact 
that a string or a wrapped membrane has mass proportional 
to its length. 
The difference of the existence of an allowed zero mode 
between a membrane and a string 
originates from the form of the BPS solutions in that 
in the former the winging number is encoded in the power, 
while in the latter its coefficient corresponds to the number 
of strings. We regard such a sharp contrast as a manifestation 
of a difference in the stability of a membrane and a string. 
Moreover we clarify how the BPS solutions we considered 
are connected with each other via the novel Higgs mechanism, 
in particular how the winding number of a membrane 
is translated into the string charge. 

It is evident that our approach provides only a sign of 
instability of a supermembrane in the static gauge and 
is far from a proof. 
If we try to prove it rigorously as in \cite{deWit:1988ct}, 
we have to regularize an action of the M2-brane 
in the static gauge which should be away from the IR fixed point. 
Therefore a kind of nonperturbative formulation of a Yang-Mills-Chern-Simons system may be necessary for it. 

As an application of our approach, it would be interesting to 
apply it to the BLG model \cite{Bagger:2006sk,Gustavsson:2007vu}. 
In fact, we have stressed the difference between 
the half-BPS solutions in the ABJM model and in the DBI action 
in whether the gauge field plays a nontrivial role. 
It would be intriguing to study what happens to a half-BPS solution 
in the BLG model. 

Another interesting question is a relation 
to stability of other classical solutions representing 
a fundamental string. For example, in \cite{Kawai:2001kg}
a fundamental string attached to a D-brane is realized 
as flux tube solutions attached to domain walls 
in various models. 
It is quite interesting to examine them from the point of view 
of world volume theories on domain walls.  

Recalling that our BPS solutions and zero modes 
are singular at the origin and hence they have ambiguities 
coming from regularizations or boundary contributions. 
In order to avoid them, it would be better to consider 
a configuration of a membrane which corresponds to 
a fundamental string connecting two D2-branes in the IIA picture 
as constructed in \cite{Hashimoto:1997px}. 
Applying our considerations to it would clarify 
issues in our approach like the tension of the classical 
solutions and so on.

\section*{Acknowledgements}
We would like to thank Koji Hashimoto, Satoshi Iso, Shoichi Kawamoto, 
Seiji Tetashima, Tamiaki Yoneya and Sen Zhang 
for useful discussions. Especially we are grateful to Yosuke Imamura for 
giving us a series of lectures on foundations of M-theory at Rikkyo University, 
which was a great help for our study. 
The work of T.K. was supported in part by Rikkyo University Special Fund for Research, 
and the work of A.M. was supported in part by a Special Postdoctoral
Researchers Program at RIKEN.
The authors would thank the Yukawa Institute for Theoretical Physics 
at Kyoto University. Discussions during the YITP workshop 
YITP-W-10-02 were useful for this work.

\appendix 

\section{Supersymmetry condition for spike D2-brane}
\label{SUSY-D2}
The supersymmetries preserved by a D2-brane are
specified by the projection condition 
$ \Gamma \epsilon = \epsilon $ 
where $\epsilon$ is a 32 component Majorana spinor 
and $\Gamma$ is given by \cite{Bergshoeff:1996tu,Bergshoeff:1997kr}
\begin{align}
&
\Gamma = 
{ 
\sqrt{ -\det g_{\mu \nu} } 
\over 
\sqrt{- \det(g_{\mu \nu} + 2 \pi \alpha' F_{\mu \nu})}
}
\sum_{n=0}^\infty
{(2 \pi \alpha')^n \over 2^n n!}
\gamma^{\mu_1 \nu_1 \ldots \mu_n \nu_n}
F_{\mu_1 \nu_1} \ldots F_{\mu_n \nu_n}
(\Gamma_{11})^n \Gamma_{(0)}, 
\label{Gamma-D2}\\
&
\Gamma_{(0)}
={1 \over 3! \sqrt{-\det g_{\mu \nu}}} 
\epsilon^{\mu \nu \rho} \gamma_{\mu \nu \rho},
\end{align}
where $\gamma_\mu$ are defined by 
$\gamma_\mu = \partial_\mu X^m \Gamma_m$ $(m=0,\ldots,9)$.
We take the static gauge $x^\mu = X^\mu$ and assume the 
following ansatz
\begin{equation}
X^i = 0 \,\,\, (i=3,\ldots, 8),\quad
X^9 = X^9(r), \quad 
F_{0r}=F_{0r}(r), \quad 
F_{0\theta} = F_{r \theta} = 0.
\end{equation}
Then only terms with $n=0,1$ are nontrivial in \eqref{Gamma-D2}
and $\Gamma$ becomes
\begin{equation}
\Gamma 
 =
{1 \over \sqrt{ 1  +(\partial_r X^9)^2 - (2 \pi \alpha' F_{0r})^2}}
\Big\{
\Gamma_{012}
+
(\cos \theta \Gamma_2 - \sin \theta \Gamma_1)
(\partial_r X^9 \Gamma_{09} 
+
2 \pi \alpha' F_{0r} \Gamma_{11}
)
\Big\}.
\end{equation}
The condition $\Gamma \epsilon = \epsilon$ is now rewritten as
\begin{equation}
(\partial_r X^9)^2=(2 \pi \alpha' F_{0r})^2, \quad
\Gamma_{012} \epsilon = \epsilon, \quad
(\partial_r X^9 \Gamma_{09} + 2 \pi \alpha' F_{0r} \Gamma_{11})^2 
\epsilon =0.
\end{equation}
Then, depending on the branch 
$\partial_r X^9 = \beta \times 2 \pi \alpha' F_{0r}$,  $\beta=\pm1$,
the D2-brane preserves the 8 supersymmetries 
specified by the following conditions
\begin{align}
\beta=+1 &: \quad 
\Gamma_{012}\epsilon = \epsilon, \quad 
\Gamma_{09(11)} \epsilon = -\epsilon,
\\
\beta=-1 &: \quad
\Gamma_{012}\epsilon = \epsilon, \quad 
\Gamma_{09(11)} \epsilon = \epsilon.
\end{align}

\section{Boundary contributions to energy}
\label{app:bc}
In this appendix we show that if we try to reproduce 
the energy, it is necessary 
to take account of boundary contributions 
in the world volume theory. 

For illustration we consider the D2-brane world volume theory 
\eqref{D2 action}, where the spike solution \eqref{spike} 
is supposed to represent a fundamental string 
ending on the D2-brane. In section \ref{D2} we show 
that fluctuations around \eqref{spike} contain a zero mode 
which takes the same form as in \eqref{spike}
\begin{align}
y=\phi=\alpha\log\frac{r}{\ell},
\label{zero mode2}
\end{align} 
where $\alpha\ll R=g_s\ell_s$. Adding this zero mode  
to the original classical solution amounts to replacing 
$R\rightarrow R+2\pi\alpha'\alpha$ as seen in \eqref{D2fluctuations} 
and hence it is expected that 
the total tension should become that 
multiplied by $(R+2\pi\alpha'\alpha)/R$. 
Thus the total energy of this configuration 
should be 
\begin{align}
E_{\rm tot}
=\frac{1}{2\pi\alpha'}\left(1+\frac{2\pi\alpha'\alpha}{R}\right)
\left(\Delta X^9_{\rm cl}+2\pi\alpha'\Delta y\right),
\label{Htot}
\end{align}
where $\Delta X^9_{\rm cl}=X^9(r=\delta)-X^9(r=\Lambda)$, 
$\Delta y=y(r=\delta)-y(r=\Lambda)$. 
However, since \eqref{zero mode2} 
is the zero mode of the Hamiltonian \eqref{D2 Hamiltonian2}, 
it does not seem to increase the energy of the classical solution. 

In order to resolve this puzzle, we take care of contributions 
from boundary terms neglected in deriving \eqref{D2 Hamiltonian2} 
from \eqref{D2 Hamiltonian}: 
\begin{align}
H_{\text{b}}=\int drd\theta
\Bigl\{
P_{A_r}\phi'+P_{A_\theta}\partial_\theta\phi
-P_{A_r}y'-P_{A_\theta}\partial_\theta y
\Bigr\}.
\label{boundary H}
\end{align}
The Gauss' law constraint \eqref{D2 gauss} tells us that 
the integrand in $H_b$ is indeed total derivative. 
If we integrate it naively, it would give rise to divergences
both in $r\rightarrow 0$ and $r\rightarrow\infty$. 
We therefore regularize it by introducing, for example, 
a cutoff at $r=\delta\ll 1$ and $r=\Lambda\gg 1$ as above, 
which in turn introduces a boundary. Thus we have to take care 
of contributions from these boundaries in order to 
calculate the energy of the configuration. 
Note that \eqref{zero mode2} of course satisfies 
the constraint \eqref{D2 gauss} 
and hence we can add \eqref{D2 gauss} to the Hamiltonian.  
In fact, the Hamiltonian in itself has ambiguity of adding a term 
proportional to \eqref{D2 gauss}. Since we are dealing 
with the quadratic Hamiltonian for fluctuations 
\eqref{D2 Hamiltonian2}, 
in the following we show that by taking account of boundary 
contributions, we can reproduce the quadratic part 
of the total energy in \eqref{Htot}, namely 
$\frac{2\pi\alpha'\alpha}{R}\Delta y$.   

At first sight, since our BPS configuration satisfies $y=\phi$, 
the boundary term does not seem to contribute which in fact 
vanishes for $y=\phi$. However, it is important to notice 
that they should obey different boundary conditions. 
Namely, $y$ has the usual Dirichlet boundary condition, 
while $\phi$ satisfies the Neumann boundary condition, 
because we have fixed the string charge\footnote
{In section \ref{D2}, we discuss this Neumann condition 
prohibits the zero mode \eqref{zero mode}, but in this appendix 
we argue that if it exists, what happens to the energy 
in order to clarify a role of boundary contributions.} 
\begin{align}
\int_{\rm S^1} F_{0r} =\text{constant},
\end{align}
which is imposed on both boundaries at $r=\Lambda$ and $r=\delta$. 
In order to switch the boundary condition, we should make 
the Legendre transformation for $\phi$ 
which amounts to adding  a boundary term 
to the Lagrangian \eqref{fluc L for D2}
\begin{align}
S_b & =-\int dt 
\bigg[
\phi(r=\Lambda) 
\int_{r=\Lambda}d\theta\, 
\frac{\delta{\cal L}}{\delta(\partial_r\phi)}
-
\phi(r=\delta)\int_{r=\delta}d\theta\, 
\frac{\delta{\cal L}}{\delta(\partial_r\phi)}
\bigg],
\label{Sb}
\end{align}
where we have assumed that $\phi$ is independent of $\theta$.
If we substitute \eqref{zero mode2} 
into \eqref{boundary H} with \eqref{Sb} taken into account, 
it is easy to see that \eqref{Sb} exactly cancels 
the contribution from $\phi$ to $H_b$ given by the first 
two terms in \eqref{boundary H}. Thus we are left with 
the boundary terms only for $y$ and by using \eqref{can Ar}, 
it is straightforward 
to check that it reproduces $\frac{2\pi\alpha'\alpha}{R}\Delta y$. 
Thus if we regularize the fluctuation Hamiltonian and take account of 
boundary contributions arising as a result of the regularization, 
we can reproduce the correct value of the quadratic part 
of the total energy\footnote
{Boundary terms which are necessary for the Legendre transformation 
to flip boundary conditions play important roles, in particular 
in the calculation of the expectation value of the Wilson loop 
in the AdS/CFT correspondence, see e.g. 
\cite{Drukker:1999zq, Drukker:2005kx, Kawamoto:2008gp}.} 
even if the BPS solution satisfies $y=\phi$.  

\


\begin{thebibliography}{99}

\bibitem{Witten:1995ex}
  E.~Witten,
  ``String theory dynamics in various dimensions,''
  Nucl.\ Phys.\  B {\bf 443}, 85 (1995)
  [arXiv:hep-th/9503124].

\bibitem{deWit:1988ct}
  B.~de Wit, M.~Luscher and H.~Nicolai,
  ``The Supermembrane Is Unstable,''
  Nucl.\ Phys.\  B {\bf 320}, 135 (1989).

\bibitem{deWit:1988ig}
  B.~de Wit, J.~Hoppe and H.~Nicolai,
  ``On the quantum mechanics of supermembranes,''
  Nucl.\ Phys.\  B {\bf 305}, 545 (1988).

\bibitem{Aharony:2008ug}
  O.~Aharony, O.~Bergman, D.~L.~Jafferis and J.~Maldacena,
  ``N=6 superconformal Chern-Simons-matter theories, M2-branes and their
  gravity duals,''
  JHEP {\bf 0810}, 091 (2008)
  [arXiv:0806.1218 [hep-th]].
  
\bibitem{Sasaki:2009ij}
  S.~Sasaki,
  ``On Non-linear Action for Gauged M2-brane,''
  JHEP {\bf 1002}, 039 (2010)
  [arXiv:0912.0903 [hep-th]].

\bibitem{Callan:1997kz}
  C.~G.~Callan and J.~M.~Maldacena,
  ``Brane dynamics from the Born-Infeld action,''
  Nucl.\ Phys.\  B {\bf 513}, 198 (1998)
  [arXiv:hep-th/9708147].

\bibitem{Gauntlett:1997ss}
  J.~P.~Gauntlett, J.~Gomis and P.~K.~Townsend,
  ``BPS bounds for worldvolume branes,''
  JHEP {\bf 9801}, 003 (1998)
  [arXiv:hep-th/9711205].

\bibitem{Rey:1998ik}
  S.~J.~Rey and J.~T.~Yee,
  ``Macroscopic strings as heavy quarks in large N gauge theory and  anti-de
  Sitter supergravity,''
  Eur.\ Phys.\ J.\  C {\bf 22}, 379 (2001)
  [arXiv:hep-th/9803001].

\bibitem{Savvidy:1999wx}
  K.~G.~Savvidy and G.~K.~Savvidy,
  ``Neumann Boundary Conditions from Born-Infeld Dynamics,''
  Nucl.\ Phys.\  B {\bf 561}, 117 (1999)
  [arXiv:hep-th/9902023].

\bibitem{Mukhi:2008ux}
  S.~Mukhi and C.~Papageorgakis,
  ``M2 to D2,''
  JHEP {\bf 0805}, 085 (2008)
  [arXiv:0803.3218 [hep-th]].

\bibitem{Ho:2008ei}
  P.~M.~Ho, Y.~Imamura and Y.~Matsuo,
  ``M2 to D2 revisited,''
  JHEP {\bf 0807}, 003 (2008)
  [arXiv:0805.1202 [hep-th]].

\bibitem{Honma:2008jd}
  Y.~Honma, S.~Iso, Y.~Sumitomo and S.~Zhang,
  ``Scaling limit of N=6 superconformal Chern-Simons theories and Lorentzian
  Bagger-Lambert theories,''
  Phys.\ Rev.\  D {\bf 78}, 105011 (2008)
  [arXiv:0806.3498 [hep-th]].

\bibitem{Pang:2008hw}
Y.~Pang and T.~Wang,
  ``From N M2's to N D2's,''
  Phys.\ Rev.\  D {\bf 78}, 125007 (2008)
  [arXiv:0807.1444 [hep-th]].

\bibitem{Fujimori:2010ec}
  T.~Fujimori, K.~Iwasaki, Y.~Kobayashi and S.~Sasaki,
  ``Classification of BPS Objects in N = 6 Chern-Simons Matter Theory,''
  arXiv:1007.1588 [hep-th].

\bibitem{Bergshoeff:1987dh}
  E.~Bergshoeff, M.~J.~Duff, C.~N.~Pope and E.~Sezgin,
  ``SUPERSYMMETRIC SUPERMEMBRANE VACUA AND SINGLETONS,''
  Phys.\ Lett.\  B {\bf 199}, 69 (1987).

\bibitem{Hashimoto:1997px}
  A.~Hashimoto,
  ``The shape of branes pulled by strings,''
  Phys.\ Rev.\  D {\bf 57}, 6441 (1998)
  [arXiv:hep-th/9711097].

\bibitem{Yoneya:1977yi}
  T.~Yoneya,
  ``Stability And Instability Of The Wu-Yang Solution Of Yang-Mills Field
  Equation,''
  Phys.\ Rev.\  D {\bf 16}, 2567 (1977).

\bibitem{Bandres:2008ry}
  M.~A.~Bandres, A.~E.~Lipstein and J.~H.~Schwarz,
  ``Studies of the ABJM Theory in a Formulation 
  with Manifest SU(4) R-Symmetry,''
  JHEP {\bf 0809}, 027 (2008)
  [arXiv:0807.0880 [hep-th]].

\bibitem{Baek:2008ws}
  J.~H.~Baek, S.~Hyun, W.~Jang and S.~H.~Yi,
  ``Membrane Dynamics in Three dimensional N=6 Supersymmetric Chern-Simons
  Theory,''
  arXiv:0812.1772 [hep-th].


\bibitem{Bergshoeff:1996tu}
  E.~Bergshoeff and P.~K.~Townsend,
  ``Super D-branes,''
  Nucl.\ Phys.\  B {\bf 490}, 145 (1997)
  [arXiv:hep-th/9611173].


\bibitem{Bergshoeff:1997kr}
  E.~Bergshoeff, R.~Kallosh, T.~Ortin and G.~Papadopoulos,
  ``kappa-symmetry, supersymmetry and intersecting branes,''
  Nucl.\ Phys.\  B {\bf 502}, 149 (1997)
  [arXiv:hep-th/9705040].

\bibitem{Drukker:1999zq}
  N.~Drukker, D.~J.~Gross and H.~Ooguri,
  ``Wilson loops and minimal surfaces,''
  Phys.\ Rev.\  D {\bf 60}, 125006 (1999)
  [arXiv:hep-th/9904191].

\bibitem{Drukker:2005kx}
  N.~Drukker and B.~Fiol,
  ``All-genus calculation of Wilson loops using D-branes,''
  JHEP {\bf 0502}, 010 (2005)
  [arXiv:hep-th/0501109].


\bibitem{Kawamoto:2008gp}
  S.~Kawamoto, T.~Kuroki and A.~Miwa,
  ``Boundary condition for D-brane from Wilson loop, and gravitational
  interpretation of eigenvalue in matrix model in AdS/CFT correspondence,''
  Phys.\ Rev.\  D {\bf 79}, 126010 (2009)
  [arXiv:0812.4229 [hep-th]].

\bibitem{Bagger:2006sk}
  J.~Bagger and N.~Lambert,
  ``Modeling multiple M2's,''
  Phys.\ Rev.\  D {\bf 75}, 045020 (2007)
  [arXiv:hep-th/0611108]; \\
  J.~Bagger and N.~Lambert,
  ``Gauge Symmetry and Supersymmetry of Multiple M2-Branes,''
  Phys.\ Rev.\  D {\bf 77}, 065008 (2008)
  [arXiv:0711.0955 [hep-th]];\\
  J.~Bagger and N.~Lambert,
  ``Comments On Multiple M2-branes,''
  JHEP {\bf 0802}, 105 (2008)
  [arXiv:0712.3738 [hep-th]].

\bibitem{Gustavsson:2007vu}
  A.~Gustavsson,
  ``Algebraic structures on parallel M2-branes,''
  Nucl.\ Phys.\  B {\bf 811}, 66 (2009)
  [arXiv:0709.1260 [hep-th]];\\
  A.~Gustavsson,
  ``Selfdual strings and loop space Nahm equations,''
  JHEP {\bf 0804}, 083 (2008)
  [arXiv:0802.3456 [hep-th]].

\bibitem{Kawai:2001kg}
  H.~Kawai and T.~Kuroki,
  ``Strings as flux tube and deconfinement on branes in gauge theories,''
  Phys.\ Lett.\  B {\bf 518}, 294 (2001)
  [arXiv:hep-th/0106103].








\end{thebibliography}
\end{document}